\documentclass{llncs}
\usepackage{fontawesome}

\usepackage{llncsdoc}
\usepackage{booktabs} 
\usepackage[colorinlistoftodos,prependcaption,textsize=tiny]{todonotes}
\usepackage{graphics,graphicx}
\graphicspath{{figures/}}
\usepackage{subfloat}
\usepackage{subfig}
\usepackage{float}

\usepackage{color-edits}
\addauthor{ef}{green}
\addauthor{ad}{blue}
\addauthor{rd}{red}

\begin{document}

\title{``Senator, We Sell Ads'': Analysis of the \\ 2016 Russian Facebook Ads Campaign}
\author{Ritam Dutt\inst{1}, Ashok Deb\inst{2} \and Emilio Ferrara\inst{2}}
\institute{Indian Institute of Technology Kharagpur, Kharagpur, 721302, India
\and
University of Southern California, Los Angeles CA 90089, USA}

\maketitle

\begin{abstract}
One of the key aspects of the United States democracy is free and fair elections that allow for a peaceful transfer of power from one President to the next. The 2016 US presidential election stands out due to suspected foreign influence before, during, and after the election. A significant portion of that suspected influence was carried out via social media. In this paper, we look specifically at 3,500 Facebook ads allegedly purchased by the Russian government. These ads were released on May 10, 2018 by the US Congress House Intelligence Committee. We analyzed the ads using natural language processing techniques to determine textual and semantic features associated with the most effective ones. We clustered the ads over time into the various campaigns and the labeled parties associated with them. We also studied the effectiveness of Ads on an individual, campaign and party basis. The most effective ads tend to have less positive sentiment, focus on past events and are more specific and personalized in nature. The more effective campaigns also show such similar characteristics. The campaigns' duration and promotion of the Ads suggest a desire to sow division rather than sway the election.
\end{abstract}

\section{Introduction}

One of the key aspects of the United States democracy is free and fair elections, unhindered by foreign influence, that allow for a peaceful transfer of power from one President to the next. Campaign Finance laws forbid foreign governments or individuals from participating or influencing the election. The 2016 US presidential election stands out not only due to its political outsider winner, Donald J. Trump, but also due to suspected foreign influence before and during the election. It is alleged that the Russian Federation operated the Main Intelligence Directorate of the General Staff, a military intelligence agency. This agency is suspected of influencing the election with resources allocated towards social media on a variety of forums.  

Corporate leadership and council from Google, Twitter, and Facebook testified on November 1, 2017 to the Senate Intelligence Committee concerning \textit{social media influence} on their platforms. While Facebook's General Council suggested that it would be difficult to verify that every ad purchased on their platform adheres to US campaign finance laws, intuitively ads purchased in Russian Rubles would be highly suspicious. 
There were approximately 3,500 ads identified by Facebook that met such criteria totaling close to \$100K and purchased between June 2015 and August 2017. The surfacing of these ads contributed to  
Facebook's CEO Mark Zuckerberg's testimony of 10-11 April 2018 to a number of Senate and House Committees.\footnote{https://www.judiciary.senate.gov/meetings/facebook-social-media-privacy-and-the-use-and-abuse-of-data}


Rep. Adam Schiff, the ranking member of the House Intelligence Committee, had voiced his opinion that the Russians had launched an `independent expenditure campaign on Trump's behalf, regardless of his involvement.\footnote{http://www.businessinsider.com/russian-facebook-ads-2016-election-trump-clinton-bernie-2017-11} However, as emphatically stated by Rob Goldman,\footnote{https://www.cnbc.com/2018/02/17/facebooks-vp-of-ads-says-russian-meddling-aimed-to-divide-us.html} the vice president of ads at Facebook, the over-arching aim of the advertisements was to bring about discord among different communities in the United States. 
In Goldman's words, ``(the ads) sought to sow division and discord'' in the political proceedings before, during, and after the 2016 US elections by leveraging the freedom of free speech and pervasive nature of social media. This statement is contradictory to the claim that the primary objective of the advertisements was to influence the effect of the 2016 elections and sway it in favor of Trump or to vilify Clinton. 
Under the direction of Democrats on the House Intelligence Committee, the Russian Facebook ads were released to the public on May 10, 2018. The main objective of this work is to apply language analysis and machine learning techniques to understand the intent of those ads by exploring their effectiveness from a campaign perspective.


\section{Related Literature}

Since the early 2000s, there has been increasing research in the new domain of \textit{computational social science} \cite{lazer2009life}. Most of the literature has focused on networked influence, information (or misinformation) diffusion, and social media association with real-world events \cite{gruzd2017social}. As it concerns our research efforts, related work focuses on social media use in politics as well as campaign detection. 

\textbf{Politics in Social Media:} Concerning divisive issue campaigns on Facebook, ongoing work has explored the organizations and objectives behind the Russian ads from a political communication standpoint. Kim \cite{kim2018stealth} stated that suspicious groups which could include foreign entities are behind many of the divisive campaigns. Additionally, approximately 18\% of the suspicious groups were Russian. The authors asserts that there are shortcomings in federal regulations and aspects of digital media that allow for anonymous groups to sow division \cite{kim2018stealth}. While Kim approaches the issue from a policy perspective, we focus more on the effectiveness and organization of the ads themselves. While the data we used in this research is specific to only the Russian Facebook ads, we present a methodology that could be extended to automatically sort any ads into their divisive campaigns. Previous work established that social media platforms were exploited during the 2016 US Presidential Election \cite{bessi2016social,allcott2017social,woolley2017computational,badawy2018analyzing}, as well as numerous other elections \cite{ratkiewicz2011detecting,metaxas2012social,howard2016bots,ferrara2017disinformation,del2017mapping,stella2018bots} and other real-world events \cite{ferrara2015manipulation,ferrara2018measuring}, by using tools like bots and trolls \cite{ferrara2016rise,davis2016botornot,ICWSM1715587}.

\textbf{Campaign Detection:} In order to combat misinformation, it is necessary to understand the characteristics that allow it to be effective \cite{mccright2017combatting}. In addition to misinformation, divisive information which creates polarized groups is counter to what the political system or a democratic nation needs to thrive \cite{sunstein2018republic}. Previous campaign detection has been focused on spam \cite{dinh2015spam} and malware \cite{saher2015malware,kruczkowski2015fp} in order to protect computer information systems. The most relevant work for campaign detection on social media is by Varol and collaborators \cite{varol2017early,ferrara2016detection}. They use supervised learning to categorize Twitter memes from millions of tweets across a series of hashtags. In comparison to that work, we focus at the microscopic level on paid Facebook ads determined to be from the same source. In addition to looking at the Russian  campaign messaging and content, we are able to factor cost and effectiveness into our analysis.  
\section{The Data}
\subsection{Collection}

The dataset comprises 3,516 advertisements with 22 variables as released by the Data for Democracy organization in csv format.\footnote{https://data.world/scottcame/us-house-psci-social-media-ads} The data was released under the direction of the Democrats on the House Intelligence Committee. The ads were released to the public on May 10, 2018. The ads were purchased in Russian Rubles during the 2016 US Presidential election and beyond from June 2015 to August 2017. In analyzing effectiveness, we only considered ads which were viewed by at least one person (impressions greater than zero). In analyzing campaigns, all ads with non-zero impressions or those which were purchased in Rubles were considered. Our dataset consists of the Russian Facebook paid ads totaling \$93K. Again, the data was initially provided by Facebook, so there is no way to independently verify its completeness and ads purchased in Rubles would be a lower bound to all ads purchased on behalf of the broader operation.  
Summary statistics of the data are shown in Table \ref{tab:data}.

\begin{table}[!h]
\centering
\caption{Summary Statistics of Russian Facebook Ad Dataset
\label{tab:data}} 
\begin{tabular}{|l|c|c|c|}
        	\hline
        		Criteria& Count& Value& Analysis \\ \hline		
			All ads&3,516&\$100K&Individual\\
			Ads with at least 1 impression&2,600&\$93.0K&Effectivness\\
			Ads with at least 1 impression AND paid in RUB &2,539&\$92.8K&Campaign \& Party\\
             \hline
\end{tabular}
\end{table}

\subsection{Preliminary Data Analysis}

\begin{figure}[htb]   
	\vspace{-1.0cm}
	\centering
    
    \subfloat[Impressions vs Ad]
    {
    \includegraphics[height=0.26\columnwidth]{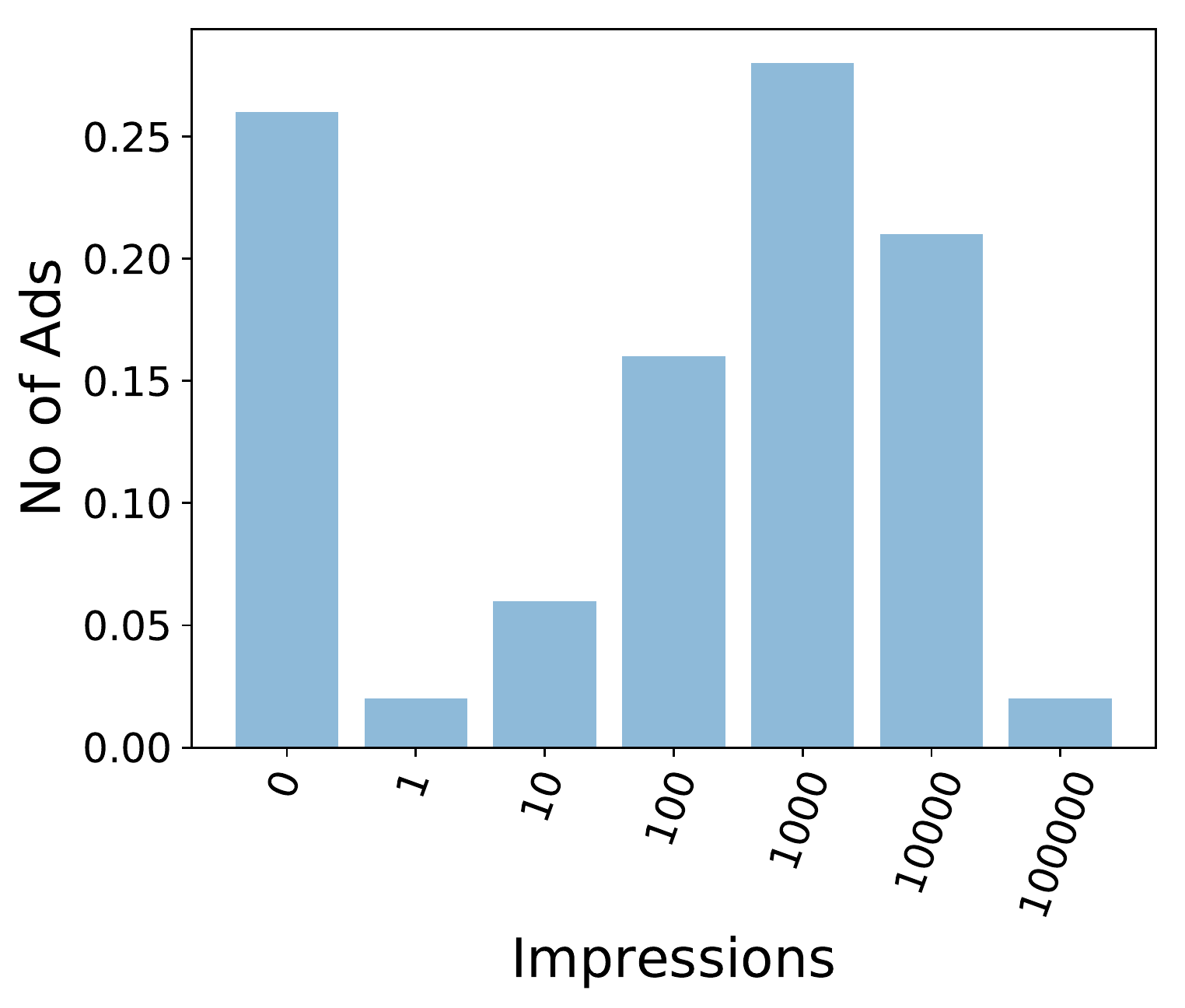}
    \label{fig:impressions-ad-dist}
	}
    \subfloat[Clicks vs Ad
    \label{fig:clicks-ad-dist}]
    {
    \includegraphics[height=0.26\columnwidth]{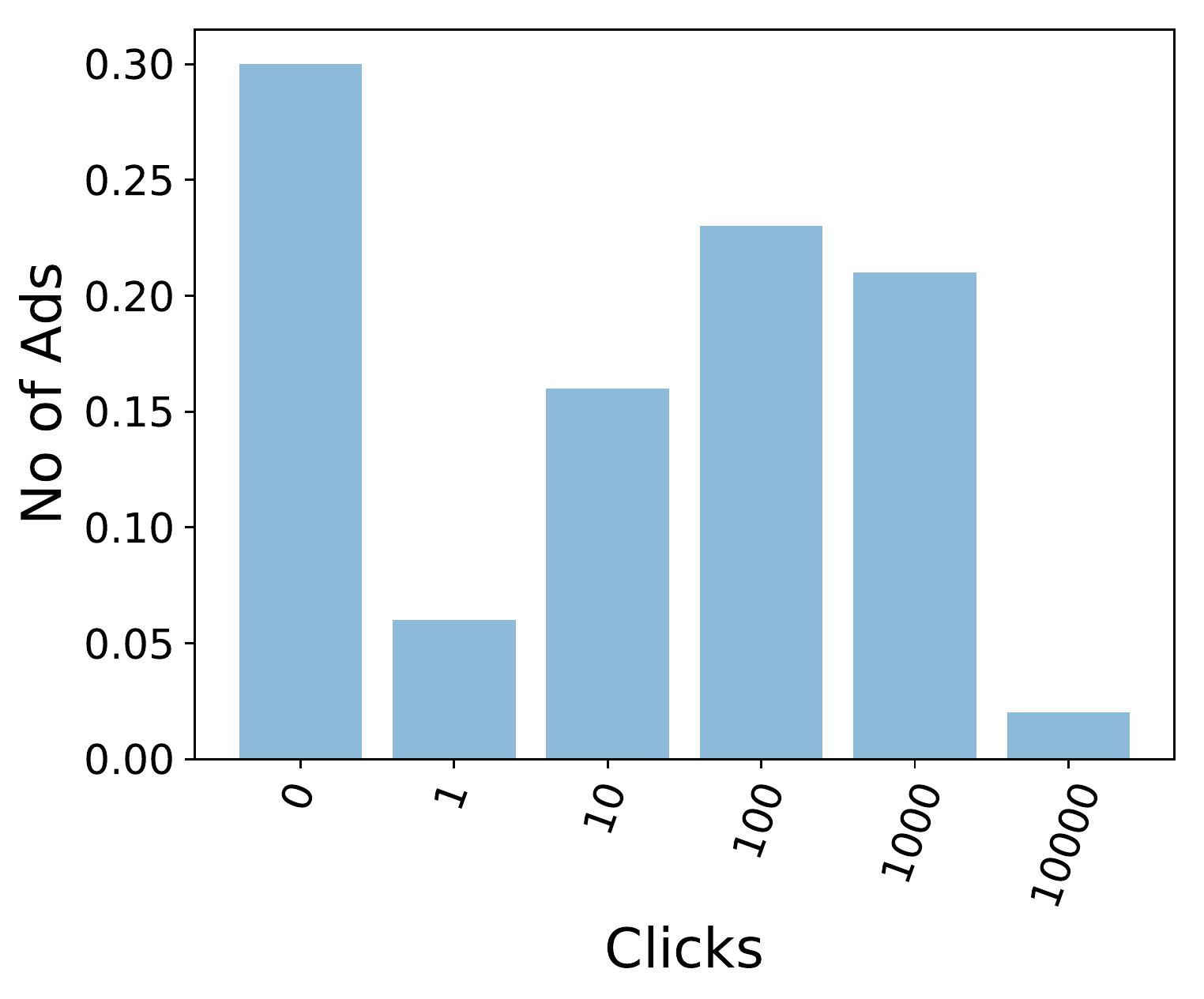}
	}
\subfloat[Target Age Group vs Ads]
    {
    \includegraphics[height=0.26\columnwidth]{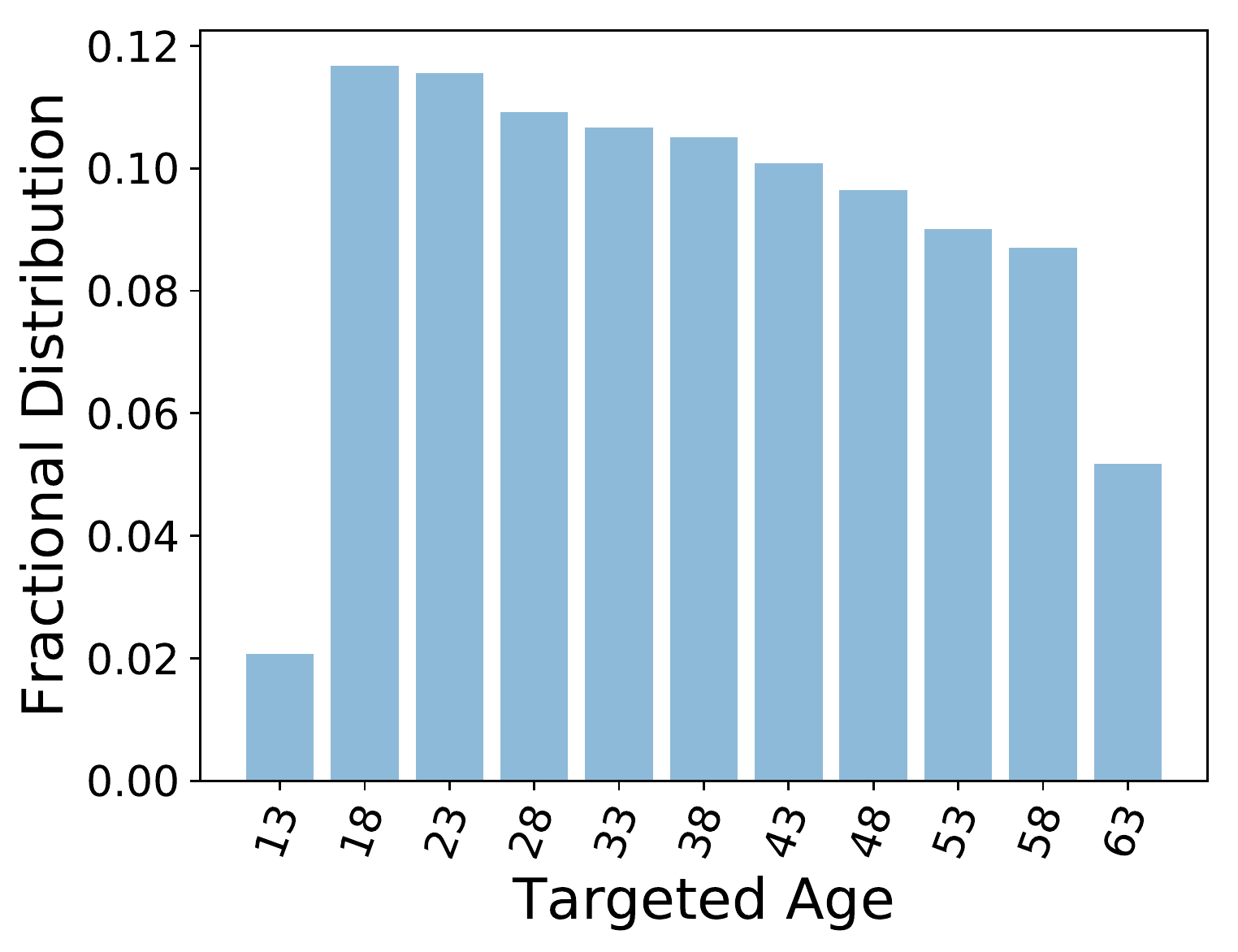}

    \label{fig:target-age-dist}
	}
    \vspace{-2mm}
    \caption{Distribution by Impressions, Clicks and Target Age Group}
    \label{fig:clks-imp-dist}
    \vspace{-7mm}
\end{figure}

\textbf{Clicks and Impression counts:}
Clicks and impressions are important metrics to understand the outreach and efficacy of the advertisement. Clicks, or link clicks, quantify the number of people who have clicked on the ad and was redirected to the particular landing-page. Impressions refer to the total number of times the advertisement has been shown. It differs from  Reach which reflects instead the  number of individual people who have seen the ad. We present the distribution of impressions and clicks for the FB ads in \ref{fig:impressions-ad-dist} and \ref{fig:clks-imp-dist}, respectively. It is clearly evident that a majority of ads have attained sufficient outreach and popularity. We observe that a huge fraction of the ads are targeted to the younger age group as seen in \ref{fig:target-age-dist}.

\section{Research Framework and Research Questions}


In this paper, we define our research framework tackling the following problem:

\vspace{0.1cm}
\noindent
\textbf{Research Question:} What features are associated with the engagement of the Russian Facebook ads and what was their impact (i.e., \textit{how effective were they}) at a campaign-wise (\textit{operational}), and on a party-wise (\textit{strategic}) basis? 

\vspace{0.1cm}
\noindent
To operationalize this question we split it into three sub-parts:

\paragraph{1. What features are associated with the engagement of the Russian Facebook ads.}
\textbf{Definition of engagement}: To quantify engagement, we estimate how likely a person would respond to an ad when it is shown to them. The metric we use is Click-Through Rate (CTR). 
We approach the problem by classifying ads which have non-zero impressions in two groups, namely \textit{more effective} and \textit{less effective} ads. The classification is done using a decision rule where the median value of the CTR across all ads is the threshold. We consider non-zero impressions only since we cannot evaluate the effectiveness of an ad that was not seen. 

We then analyze the stylistic and textual features of the Ads between the two categories, using different natural language processing techniques. The features include sentiment, emotion, structural content, parts of speech distribution, named entity distribution, and linguistic categories. We note those features which show significant differences across categories. 
\vspace{-0.1cm}
\paragraph{2. What was the Ads impact (effectiveness) at a campaign-wise (operational) level?}
\textbf{Definition of effectiveness}: At the campaign level we define effectiveness as the audience reach efficacy. The metric we use is Cost Per Thousand Impressions (CPM) and Cost Per Click (CPC) (explained in methodology). We approach this question by clustering the ads into the various campaigns and using CPM to determine the most and least effective campaigns as well as any insights from the associated features mentioned in the first sub-part.
\vspace{-0.1cm}
\paragraph{3. What was the Ads impact (effectiveness) at a party-wise (strategic) level?}
\textbf{Definition of effectiveness}: We define the effectiveness at the party level by observing significant differences in terms of CTR, CPM and total cost between the parties. We create parties by manually labeling the ads into Democratic (Blue), Republican (Red) and Neutral (Green). We exclude the Neutral campaigns and assess the effectiveness of the Blue and Red parties and report any significant findings from a feature-wise perspective. 

It is notable to mention our assumption that all of these ads within the campaigns and parties were generated by the same alleged organization in Russia. 

\section{Methodology}

\subsection{Features of Effective Ads}
The effectiveness of ads at the individual level is measured using CTR. 

\textbf{Click-Through Rate (CTR)}. CTR of a particular advertisement is the ratio of clicks to impression for the ad expressed as a percentage. CTR reflects the creativity and compelling nature of the advertisement \cite{dave2010learning}. 
  \begin{equation}
  CTR=\frac{\#Clicks}{\#Impressions}*100(\%)
  \end{equation}
\vspace{0.2cm}


\noindent The stylistic and textual features associated with the ads we analyzed are:


\textbf{Sentiment Analysis:} Sentiment analysis helps to identify the attitude of the text and gauge whether it is more positive, negative or neutral.  Based on the comparative analysis of in \cite{sentibench}, we utilized 2 methods to determine sentiment on the Ad text to obtain the overall compounded sentiment score of the Ad. VADER: Valence Aware Dictionary for Sentiment Reasoning \cite{hutto2014vader} is a rule-based sentiment model that has both a dictionary and associated intensity measures. Its dictionary has been tuned for micro-blog like contexts. We also observe the categories corresponding to positive and negative emotions by performing LIWC \cite{pennebaker2001linguistic} analysis on the \textit{Ad Text}. 


\textbf{Emotion Analysis:} We leverage the NRC lexicon of \cite{NRC-lexicon} to calculate the average number of words corresponding to an emotion per advertisement. The associated 8 emotions include anger, anticipation, joy, fear, trust, disgust, sadness and surprise.

\begin{figure}
\centering
  \includegraphics[width=65mm]{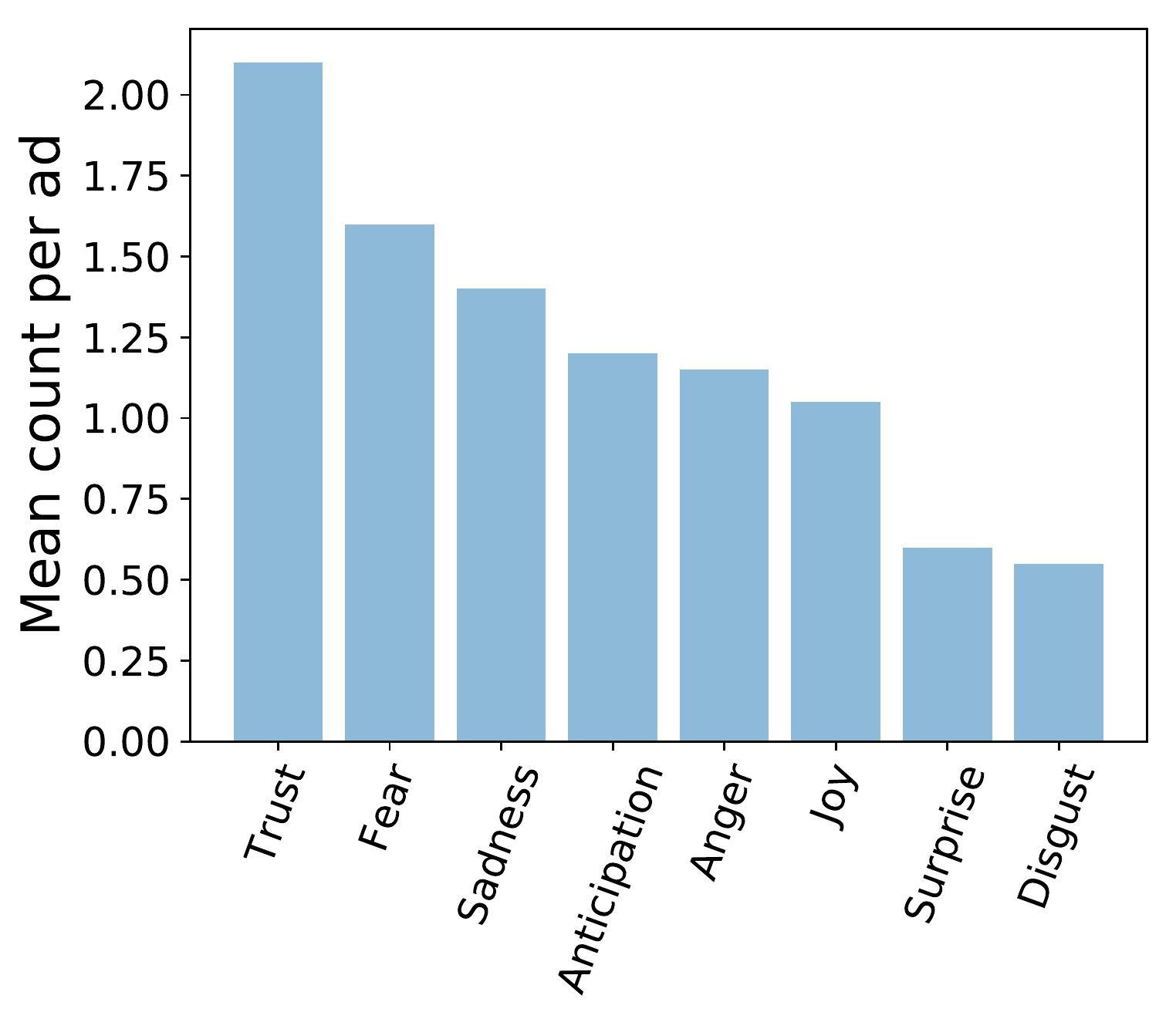}
  \caption{Emotion word counts}
  \label{fig:emotion}
\end{figure}

\textbf{Structural Content:} The structural content of the text refers to the \textit{distribution of sentences and words} per advertisement. An ad's efficiency often correlates with the amount of textual content \cite{structalfec}.

\textbf{POS-TAG distribution:} We employed the inbuilt Part-of-Speech (POS) tagger of NLTK \cite{POSTAG} and the Penn Tree Bank \footnote{https://www.ling.upenn.edu/courses/Fall\_2003/ling001/penn\_treebank\_pos.html} to observe the distribution of different POS (Parts of Speech) TAGS in the advertisement texts. 

\textbf{Named Entity Recognition (NER) distribution:} The high proportion of proper nouns from POS TAG analysis signifies that the ads cater more to real-world events. Consequently, we also inspected the distribution of different named entities using the Perceptron-based NER of Stanford CoreNLP \cite{CoreNLP} pertaining to "PERSON", "ORGANIZATION" and "LOCATION". 

\textbf{Linguistic Inquiry and Word Count (LIWC) Analysis:} LIWC \cite{pennebaker2001linguistic} computes the proportion of words in psychologically meaningful categories for the analyzed text which we leverage to discover different linguistic and cognitive dimensions.

\vspace{-0.3cm}
\subsection{Campaign-Level Analysis}
We leverage different methods to cluster Ads into non-overlapping campaigns:

\textbf{LDA Topic extraction}:
We implemented LDA \cite{LDA} (Latent Dirichlet Allocation) using the in-built gensim model of \cite{gensim} on the corpus of advertisement text to obtain a list of 50 topics. However these topics had several overlapping words and dealt with racism, gun-control or police accountability. It also failed to capture broad topics like homosexuality or immigration.  

\textbf{Key-word/ Key-phrase extraction}: 
We also employ RAKE (Rapid Automatic Keyword Extraction) \cite{RAKE}, an unsupervised, domain-independent and language-independent technique to extract keywords from the advertisement texts. This methodology captures niche topics since it observes each document individually.

However, the above methods suffer two shortcomings. Firstly, they did not take into account the Ad's images which serves the purpose of propagating ideas mentioned in the advertisement. Secondly, these techniques do not incorporate the context associated outside of the text. For example, the ad text, "The blue gang is free to do whatever they want" clearly refers to police brutality, often misdirected at African-Americans, but it is impossible to decipher from the text alone. Hence we resort to a semi-automated network clustering technique to identify campaigns as described below. 

\textbf{Network-based Clustering:} Some of the advertisements have a meta-data field labeled Interests which corresponds to topics. We represent each topic as a vector, obtained by the FastText technique of \cite{fasttext}. We compute similarity between the topics using the given equation:

  \begin{equation}
  sim(T_{i}, T_{j})= \alpha (\vec T_{i} \cdot \vec T_{j})+ (1-\alpha) \frac{|Ads(T_{i})\cap Ads(T_{j})|}{min(|Ads(T_{i})|,|Ads(T_{j})|)}
  \end{equation}

wherein, $T_{i}$ and $T_{j}$ represent two arbitrary topics, $\vec T_{i}$ and $\vec T_{j}$ denote their vector representation and $Ads(T_{i})$ enlists the ads which have $T_{i}$ as a topic.
The first part of the equation ($\vec T_{i} \cdot \vec T_{j}$) simply computes the cosine similarity score of the topic vectors, while the second part calculates the overlap coefficient similarity between two topics.  While $\alpha$  $\epsilon$ [0,1] determines the trade-off between the two similarity scores.

Each topic is then represented as a node in an undirected graph with edges representing the similarity between two topics. We binarize the graph by retaining only the edges above a certain threshold $\beta $ and cluster it. We experimented with different values of $\alpha$, $\beta$ and different algorithms and experimentally verified that the Louvain algorithm  \cite{blondel2008fast} with a threshold of 0.9 for both $\alpha $ and $\beta$ gave the best results with 9 non-overlapping campaigns. A change in $\alpha$ and $\beta$ values drastically altered the number of communities, ranging from 2-3 on one extreme to 40-50 in another. Likewise, ML based unsupervised clustering techniques like KMeans or Spectral Clustering were unable to incorporate the overlap coefficient similarity and hence showed poorer performance.

Thus each topic belongs to one of the initial 9 campaigns. Since an ad can contain several topics, they can belong to different campaigns, we assign them to the campaign that with the most number of topics, breaking ties arbitrarily. 

We then manually inspected the rest of the ads and assigned those which did not have the Interests field to one of the 9 campaigns. Sometimes, we had to create new campaigns since the particular Ad did not conform with any of the previous ones. It was necessary to break up large clusters which had similar notions (police brutality, racism and Black Lives Matter) into different campaigns. Eventually, that yielded the final 21 campaigns as demonstrated in Table \ref{tab:campaign_parties}.

\vspace{0.1cm}
\noindent The effectiveness of ads at the campaign level is measured via CPM and CPC.  

\textbf{Cost Per Thousand Impressions (CPM):} CPM for an ad is simply the amount of Rubles spent to reach a mile (thousand impressions). CPM is primarily determined by the target audience \cite{baltas2003determinants}.

\begin{equation}
  CPM=\frac{AdCost(RUB)}{\#Impressions}*1000
\end{equation}

\textbf{Cost Per Click (CPC):} CPC for an Ad is the amount of Rubles required to receive a click. CPC reflects the traffic generated by the ad to the landing page \cite{baltas2003determinants}. 

\begin{equation}
  CPC=\frac{AdCost(RUB)}{\#Clicks}
\end{equation}

A campaign's effectiveness is usually measured by a \textit{low CPC } value because it implies that the amount of Rubles required to get an audience's response is also less. However a low CPM is sometimes essential if one wishes to target a particular audience and optimize the overall cost of the campaign. If an ad itself has a high CTR, purchasing Ads using CPM may be a better strategy. 

\noindent The stylistic features analyzed are consistent with those outlined in Section 5.1. 


\subsection{Party Clustering}
The campaigns are manually assigned to parties as stated in Table \ref{tab:campaign_parties}.


\begin{table}[htb]
\centering
\resizebox{1.0\textwidth}{!}{%
\begin{tabular}{|c|c|c|}
\hline
Campaign & Definition & Party\\ \hline
Police Brutality & Injustice meted out to the Blacks by the Police & Democrat\\
Entertainment & Multi-media sources of entertainment (memes, songs, videos)& None\\
Prison & Prison reforms against mandatory sentences, prison privatization & Democrat\\
Racism & Acts of racism harbored against any racial minority in America & Democrat\\
LGBT& Rights and dignities for the LGBT people & Democrat\\
Black Lives Matter& Incarceration, shooting or other acts of cruelty against Blacks & Democrat\\
Conservative& Ideals of patriotism, preserving heritage and Republican advocacy& Republican\\
Anti-immigration& Preventing illegal immigration across the US borders& Republican\\
Veterans& Support for the hapless/ crippled veterans of war & Republican\\
2nd Amendment & Supporting the right to bear arms and guns& Republican \\
All Lives Matter& Counter to the Black Lives Matter
& Republican\\
Anti-war & Opposition of wars and acts of aggression against the Middle East & Democrat\\
Texas&A medley of Conservative Ads specifically leaned towards Texas.&Republican\\
Islam& Against Islamophobia and support for the Muslims in the US&
Democrat\\
Immigration & Support the immigration of other nationalities into America & Democrat \\
Liberalism & In support of the various liberal reforms by the Blue part & Democrat\\
Religious & Support for the conservative Christians in the US& Republican\\
Hispanic & Support for the Hispanic/ Latino community in the US& Democrat\\
Anti-Islam& Messaging against the acceptance of Muslims in US& Republican\\
Native & Support for the Native American Indians and their community& Democrat\\
Self-Defense& Focused on martial arts training for anti-police brutality  & Democrat\\ \hline

\end{tabular}}
\caption{Campaigns identified in the dataset and parties associated with them.}
\label{tab:campaign_parties}
\vspace*{-8mm}
\end{table}

\noindent The effectiveness of ads at the party level is measured using CPC and CPM, in a fashion similar to the campaign-wise analysis. 


\section{Results}

\subsection{Ad Effectiveness in Aggregate}

The calculated median CTR value of the advertisements is 10.24. Consequently, we categorize the ads as more or less effective if the CTR value is greater or lesser than 10.243 respectively.
We present the significant semantic and textual features here. In all cases, significant difference refers to a \textit{p-value} of $\leq 0.001$.


\textbf{Sentiment Analysis:} 
We observe that the overall compounded score is significantly lower for the more effective ads than those in the less effective ads, implying that the former ads tend to be less positive. Surprisingly, there is no significant difference between the distribution of negative sentiments.

\textbf{Emotion Analysis:} 
None of the 8 emotions showed any significance difference across the two categories, except surprise which demonstrated mild significance (\textit{p-value} $\leq 0.05$).

\textbf{Structural content:} The distribution of sentences and words per advertisement do not vary significantly across the two categories.


\textbf{POS-TAG distribution:}
We observe that adverbs (RB) and past tense verbs (VBD) occur more frequently in the more effective ads. This implies that more effective ads tend to refer to past events more frequently while the pronounced usage of adverbs implies that actions are explained in detail. However, the proportion of nouns across advertisements is very high, with NN (common nouns, singular) and NNP (proper nouns, singular) accounting for 6.32 and 5.89 words per advertisement respectively.

\textbf{NER distribution:} 
The NER analysis revealed that the category  "PERSON" occurred in significantly higher proportion among the more effective ads.

\textbf{LIWC Analysis:} 
Only the most significant LIWC categories have been taken to account here. 

\textit{Personalization:} Categories belonging to \textit{SheHe} and \textit{Ipron (personal pronouns)} are higher in more effective ads, while those belonging to \textit{We} and \textit{Friends are lower} in the more effective category. This indicates the more effective ads are more personalized or cater to the individuals rather than the communities.

\textbf{Religion and Money: }\textit{Religion and Money}  occur in lower proportions in the more effective ads than the less effective ones. This shows that religious or financial divide are not as successful to ensure engagement.



We now present the differences in Table \ref{tab:sig-differences}. The columns corresponding to Less Effective(Mean) and More Effective(Mean) specify the mean value of the distribution for the  categories. The Mean Diff column is simply computed
\begin{equation}
Mean Diff =\frac{High(Mean)-Low(Mean)}{Low(Mean)}*100\%
\end{equation}
The stars beside a category name correspond to the level of significance as indicated by the \textit{p-value}.

\begin{table}[htb]
\small 
\centering
\begin{tabular}{|c|c|c|c|c|}
\hline

Category & Less Effective& More Effective  &  Mean Diff & T-value \\ \hline
Compounded sentiment***&  0.139& 0.048& -65.699& 3.687\\
Positive sentiment****& 0.19& 0.158& -16.61& 
4.414\\
Negative sentiment& 0.097& 0.089& -8.243 & -6.281\\ \hline
Surprise**& 0.503&0.64& 27.4& -2.836\\
Anger & 1.061&	1.174&	10.648& -1.37\\ \hline

\#Sentences& 3.768&	3.733&	-0.939& 0.232\\
\#Words& 48.008& 52.648& 9.666& -1.873\\ \hline

RB (Adverb)**** & 1.454&	2.016&	38.698& -4.85\\
VBD (Verb, past)**** & 0.896&	1.414&	57.834& -4.927\\
NN (Common nouns)& 6.296&	6.555&	4.111& 
-0.776\\
NNP (Proper nouns) & 6.361&	6.044&	-4.983& 0.93\\ \hline

PERSON*** & 0.017	&0.028&	61.658& -3.209\\
LOCATION* & 0.012	&0.009&	-21.195& 2.127\\
\hline

Ipron****& 0.028 &	0.04&	43.807& -5.973\\
We****& 0.033&	0.017&	-48.568& 8.157\\
SheHe****&0.005&0.013&	148.432& -6.904\\
Friends****&0.004&	0.001&	-64.348&4.588\\
Money****& 0.01&0.005&	-48.167	& 5.423\\
Religion****&0.01&	0.004&	-61.087	& 4.424\\
\hline

\end{tabular}
\caption{Average values between more effective and less effective.  Significance of the feature as denoted by *,**,***,**** correspond to \textit{p-values} less than 0.05,0.1,0.001 and 0.0001 respectively.}
\label{tab:sig-differences}
\vspace{-10mm}
\end{table}

\subsection{Campaign-wise Analysis}

We present the statistics of the different campaigns in Table \ref{tab:camp-stats} which are arranged in decreasing order of their effectiveness and thus in increasing order of CPM. We demarcate the campaigns into more and less effective based on the median value of the CPM (A more effective campaign has a CPM score less than 277.57).

\begin{table}
\centering
\resizebox{1\textwidth}{!}{%
\begin{tabular}{|c|r|r|c|r|r|c|c|c|}
\hline
Topics&Cost in RUB &Cost in USD&Frequency&Impressions&Clicks& CPM&CPC&CTR\\ \hline
\textcolor{blue}{Hispanic}&164,146.40&2,628.05&186&5,943,904&713,804&27.62&0.23&12.01\\
\textcolor{blue}
{Immigration}&2,971.30&47.76&10&74,344&10,762&39.97&0.28&14.48\\
\textcolor{red}
{All Lives Matter}&150,372.36&2,368.50&11&1,890,020&82,779&79.56&1.82&4.38\\
\textcolor{blue}
{Black Lives Matter}&
1,807,407.97&	28,631.85&	1206&	19,273,576&	1,856,476&	93.78&	0.97&	9.63\\

\textcolor{green}
{Entertainment}&90,188.75&1,407.42&159&885,273&87,956&101.88&1.03&9.94\\
\textcolor{blue}
{Racism}
&	237,900.47&	3,677.33&	125&	1,364,627&	82,168&	174.33&	2.9&	6.02\\
\textcolor{blue}
{Native}&9,397.14&160.94&12&47,428&5,355&198.13&1.75&11.29\\
\textcolor{red}
{Religious}&212,647.46&3,543.32&21&1,032,898&78,669&205.87&2.7&7.62\\
\textcolor{red}
{2nd Amendment}&234,324.96&3,833.16&50&1,119,281&87,986&209.35&2.66&7.86\\
\textcolor{blue}
{Police Brutality}&563,945.02&8,873.97&194&2,535,621&207,233&222.41&2.72&8.17\\
\textcolor{red}
{Veteran}&
220,615.91&	3,468.31&	97&	794,826&	59,925&	277.57&	3.68&	7.54\\
\textcolor{red}
{Conservative}&
831,223.67&	13,600.98&	116&	2,773,169&	213,894&	299.74&	3.89&	7.71\\
\textcolor{red}
{Anti-Islam}&	4,385.58&	69.64&	3&	13,949&	2,725&	314.4&	1.61&	19.54\\
\textcolor{blue}
{LGBT}&303,738.01&4,796.96&95&887,058&82,217&342.41&3.69&9.27\\
\textcolor{blue}
{Anti-war}&27,469.85&444.45&15&75,517&6,980&363.76&3.94&9.24\\
\textcolor{blue}
{Islam}&271,567.36&4,271.96&56&581,392&22,033&467.1&12.33&3.79\\
\textcolor{blue}
{Liberalism}&87,405.43&1,387.71&33&177,089&15,542&493.57&5.62&8.78\\
\textcolor{red}
{Texas}&295,043.68&4,698.09&35&589,409&51,400&500.58&5.74&8.72\\
\textcolor{blue}
{Prison}&13,552.58&215.30&19&25,954&1,981&522.18&6.84&7.63\\
\textcolor{blue}
{Self-defense}&30,982.02&518.22&25&53,712&2,136&576.82&14.5&3.98\\
\textcolor{red}
{Anti-Immigration}&	289,898.95&	4,432.61&	71&	419,380&	57,865&	691.26&	5.01&	13.8\\

\hline
\end{tabular}}
\caption{Statistics of the campaign arranged in decreased order of effectiveness. }
\label{tab:camp-stats}
\vspace{-10mm}
\end{table}

	


We note the following stylistic differences between the more effective and less effective campaigns. 

\textbf{Sentiment Analysis:} The compounded sentiment score is significantly lower for the more effective campaigns since those campaigns involve serious topics like police brutality, racism, etc.

\textbf{Emotion Analysis:} All 8 emotions, barring surprise, are observed to be significantly pronounced in the more effective campaigns. We hypothesize that ads evoking emotions are likely to be shared more and hence the impressions increase for the ad, thereby decreasing the potential CPM. 

\textbf{Structural Analysis:} Surprisingly, we note that ads in the more effective campaigns tend to be of shorter length, i.e more concise. 

\textbf{POS-TAG distribution:} POS corresponding to adverbs (RB), plural nouns (NNS, NNPS) and verbs (VB) occur more frequently in the less effective campaigns, the significance of which is unknown. 

\textbf{Named Entity distribution:}
Named entity mentions corresponding to 'PERSON' is significantly higher in the more effective campaigns while 'LOCATION' is higher in the less effective ones. This finding is attributed to disproportionate large mentions of victims of racial prejudice in the more effective campaigns. Likewise, the less effective campaigns include Texas, Anti-Immigration to US, Veterans, etc which directly reference America. 

\textbf{LIWC Analysis:}
In the category of \textit{Religion}, the less effective campaigns have a higher proportion of ads associated with Islam. This conforms the analysis at the individual ad level that religious ads  are less effective. As for \textit{Associativity}, the more effective campaigns are also individualistic/personal as opposed to community-driven. This finding is substantiated by the significantly high frequency of \textit{I} and \textit{We} categories respectively in the more and less effective campaigns.

We also observe that the individual CPM and cost spent on an ad is significantly lower for the more effective campaigns than the less effective ones. Likewise, the number of 
clicks and CTR of an individual Ad is significantly higher for the more effective campaigns. Thus, more effective ads do contribute to effective campaigns, although the effectiveness metrics themselves are different for the parties and campaigns. 
\vspace{-0.2cm}

\subsection{Party-wise Analysis}
We present a statistical overview of the ads of the two parties in Table \ref{tab:part-stats}. 
\vspace{-0.5cm}

\begin{table}[htb]
\centering
\small
\resizebox{1.0\textwidth}{!}{%
\begin{tabular}{|c|c|c|c|c|c|c|c|c|}
\hline
Party&\# Ads&Cost&Cost&Clicks&Impressions&CPM(RUBs)&CPC(RUBs)&CTR\\ \hline
Democrat&1,976&3.5MRUB&\$55.6K&2,995K&31.0M&113.42
&1.17&9.69 \\
Republican&404&2.2MRUB&\$36.0K&647K&8.7M&259.30&3.52&7.36\\
\hline
\end{tabular}}
\caption{Performance of the two parties.}
\label{tab:part-stats}
\vspace{-1mm}
\end{table}
\vspace{-0.8cm}

Although there is no significant difference in the distribution of clicks and impressions between the Ads of the two parties, the Democratic party had significantly higher CTR and lower CPM values. This implies that the Democratic party was more effective amongst the two parties.  

\begin{figure}[H] 
\centering
  \includegraphics[width=\columnwidth]{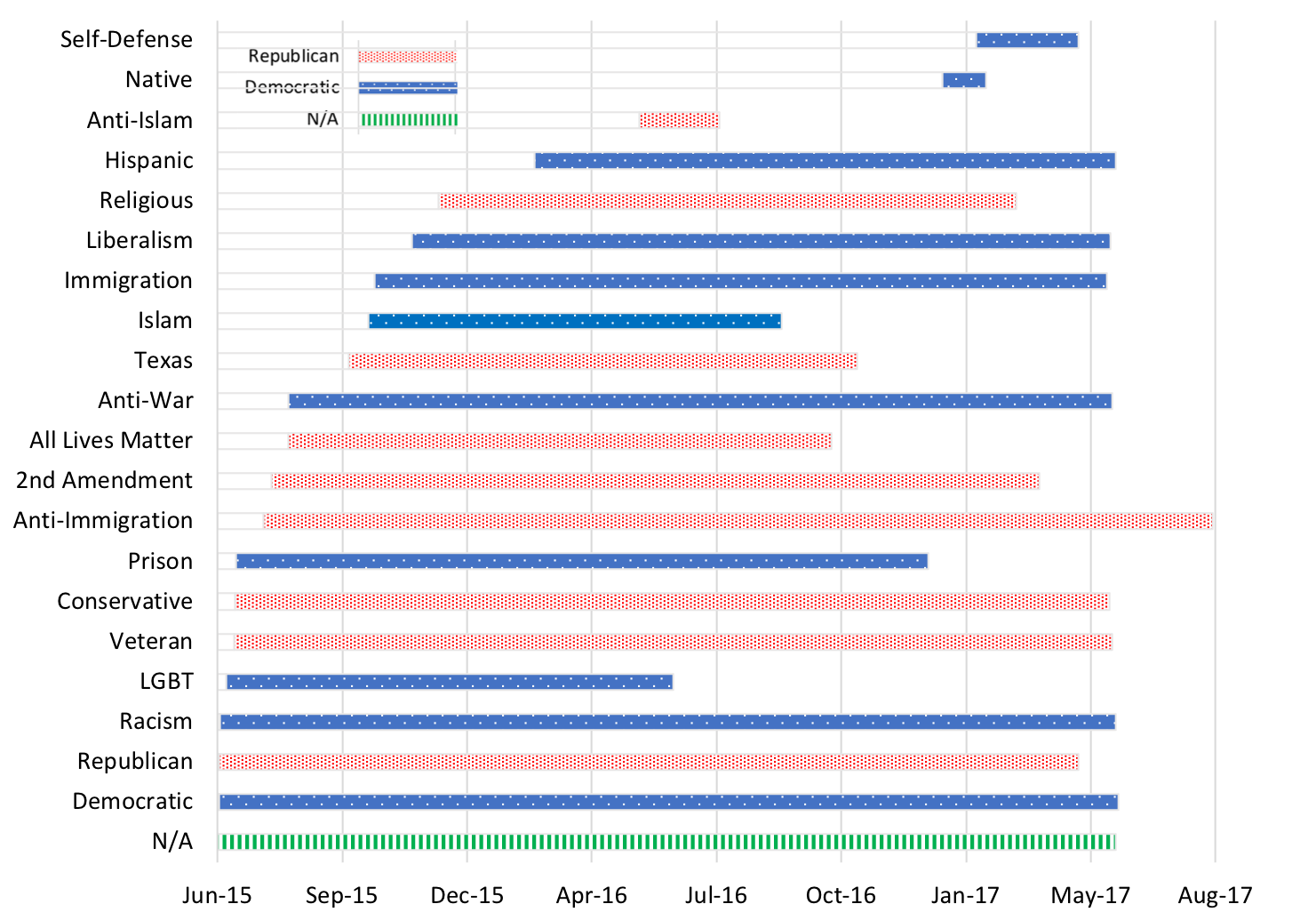}
  \caption{Timeline of Campaigns}
  \label{fig:camptime}
\end{figure}

However, there was also an active involvement in propagating the Republican Ads as well. This is evident from Table \ref{tab:part-stats} which highlights that the disproportionate high amount spent for the Republican Ads (38.87\%) despite their low frequency (17.10\%). Moreover, adjudging from the campaign's time-line in Figure \ref{fig:camptime} Republican Ads occurred for a longer duration. 

Finally, the campaigns of the two parties mostly dealt with conflicting or contradictory ideals (Anti-Islam/Islam, Anti-Immigration/Immigration, All Lives Matter/Black Lives Matter). This strongly suggests desire to sow discord.

\noindent We now present the semantic and textual differences between the two parties.

\textbf{Sentiment Analysis:} The compounded sentiment score is significantly lower for the Democratic party since a majority of the Democratic ads pertain to serious topics like police brutality, racial tension, anti-war, etc.

\textbf{Emotion Analysis:} The emotion corresponding to sadness is significantly more pronounced in the Democratic ads due to the above reason. 

\textbf{Structural Analysis:} There was no significant difference in the average distribution of words and sentences between the two parties.


\textbf{POS-TAG distribution:} Surprisingly, plural nouns (both common and proper nouns) occur more frequently in the ads of the Republican party. Adverbs and comparative adjectives are also more prevalent in the Republican ads.

\textbf{Named Entity distribution:} The fraction of named entities corresponding to \textit{Person} is higher in Democratic ads while those corresponding to \textit{Location} is higher in Republican ads. This happens since the Democratic ads mention the names of victims of racial prejudice like Tamir Rice and Eric Garner. Republican ads of patriotism, veterans, and 2nd Amendment indirectly referenced America. 

\textbf{LIWC Analysis:} 
The category \textit{We} is more significantly pronounced in the Republican party than the Democratic party which might indicate a closer community or inclusiveness. This may be appealing to the target's sense of belonging. 

\begin{figure}[H]
  \centering
  \label{figur}

  \subfloat[\textit{Democratic} highest clicks (56K) and impressions (968K)]{\label{figur:1}\includegraphics[width=60mm]{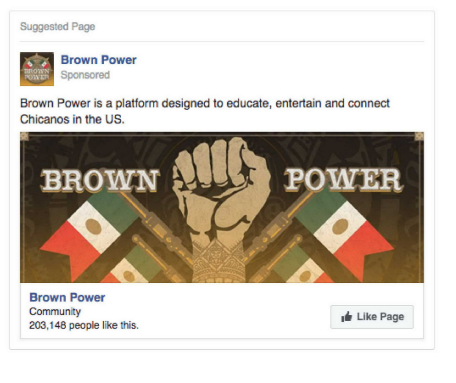}}
  \subfloat[\textit{Democratic} highest CTR (84.42\%)]{\label{figur:2}\includegraphics[width=60mm]{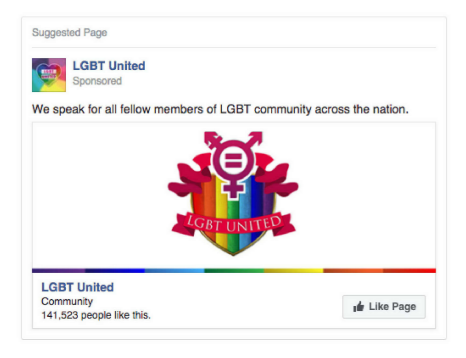}}
  \\
  \subfloat[\textit{Democratic} highest cost (\$1,200)]{\label{figur:3}\includegraphics[width=60mm]{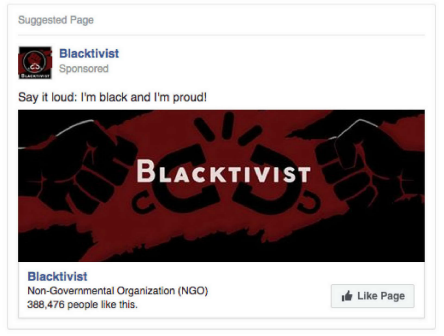}}
  \subfloat[\textit{Republican} highest clicks (73K) and impressions (1.33M)]{\label{figur:4}\includegraphics[width=60mm]{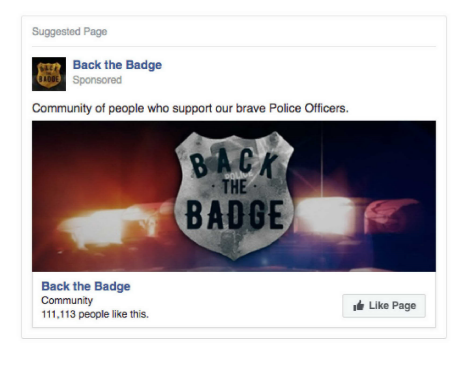}}
  \\
  \subfloat[\textit{Republican} highest CTR(28.16\%)]{\label{figur:5}\includegraphics[width=60mm]{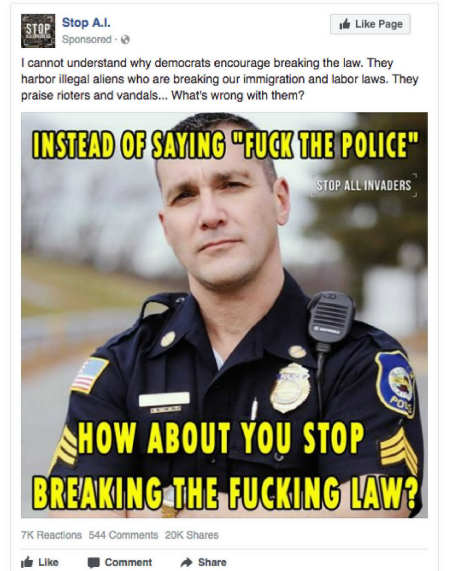}}
  \subfloat[\textit{Republican} highest cost (\$5,317)]{\label{figur:6}\includegraphics[width=60mm]{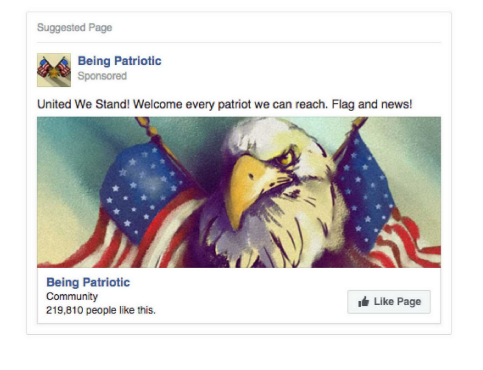}}
\caption{Best performing ads for each party}
\end{figure}

\begin{figure}[H]   
	\vspace{-1.0cm}
	\centering
    
    \subfloat[Ad Distribution by Cost]
    {
    \includegraphics[width=0.75\textwidth]{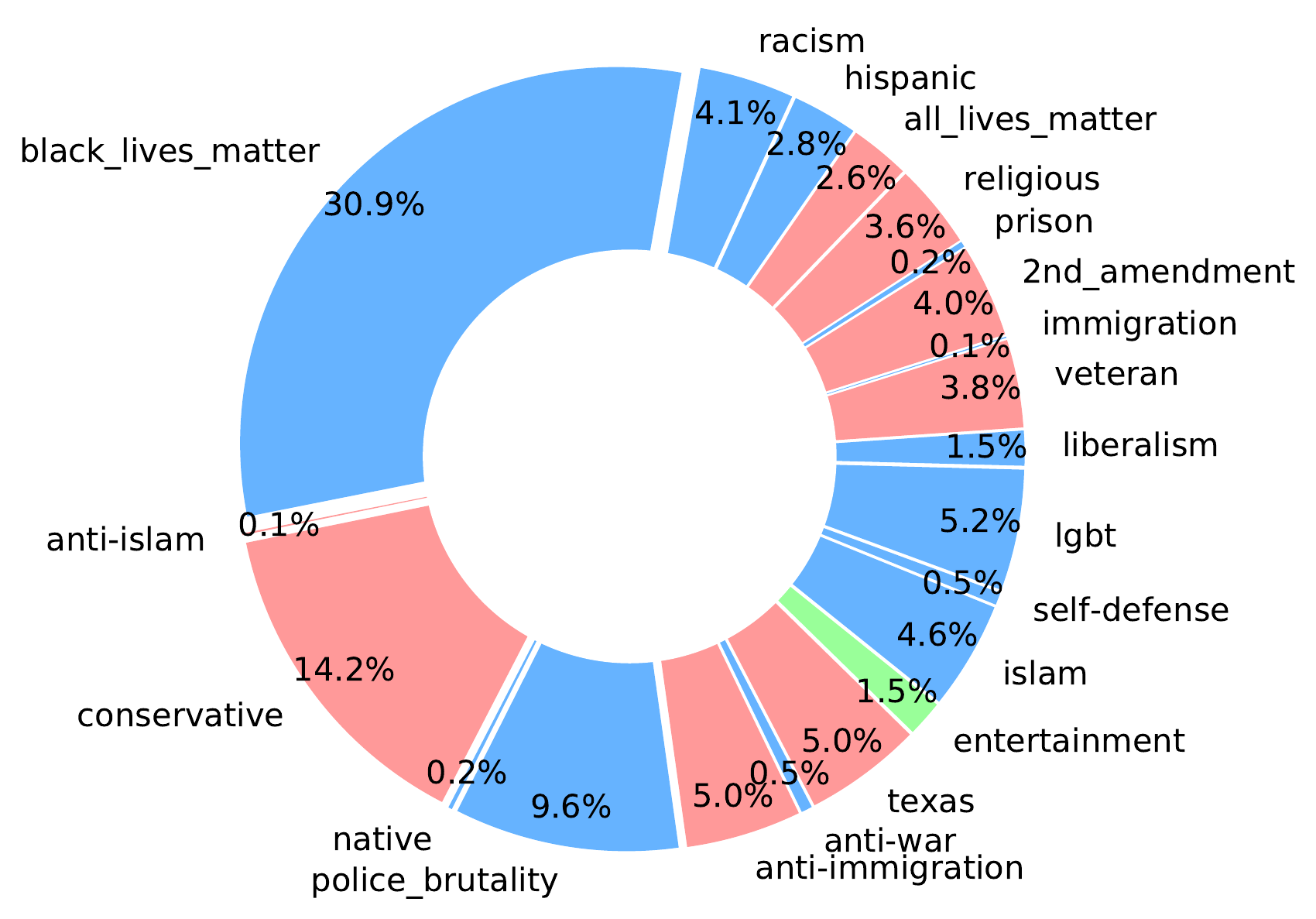}
    \label{fig:camp_cost}
	}
	
    \subfloat[Ad Distribution by Count]{
    \includegraphics[width=0.75\textwidth]{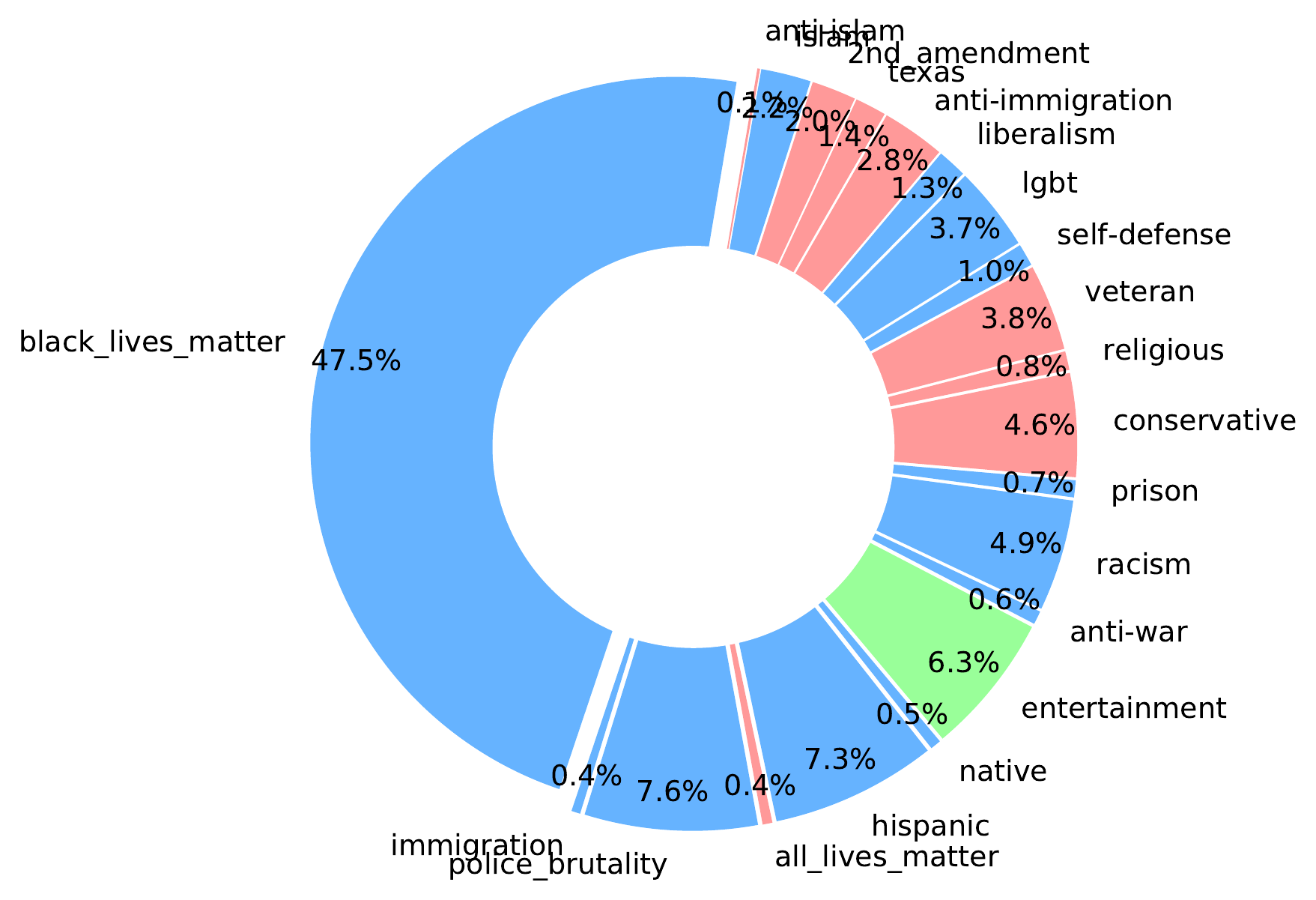}
	\label{fig:camp_count}
	}
    \vspace{-2mm}
    \caption{Distribution by Impressions, Clicks and Target Age Group}
    \label{fig:pie_chart}
    \vspace{-7mm}
\end{figure}

\section{Conclusions and Future work}
In this paper we characterized the Russian Facebook influence operation that occurred before, during, and after the 2016 US presidential election. We focused on 3,500 ads allegedly purchased by the Russian government, highlighting their features and studying their effectiveness. The most effective ads tend to have less positive sentiment, focused on past events and are more specific and personalized in nature. A similar observation holds true for the more effective campaigns. The ads were predominately biased toward the Democratic party as opposed to the Republican party in terms of frequency and effectiveness. Nevertheless the campaigns' duration and promotion of the Republican Ads do hint at the efforts of the Russians to cause divide along racial, religious and political ideologies. Areas for future work include exploring other platforms and similar operations carried therein. For example, we would like to investigate the connection to Russian troll accounts identified on Twitter, and conduct campaign analysis to determine the effectiveness of such operations across various platforms. 

\bigskip

 \noindent\textbf{Acknowledgement}. EF is grateful to AFOSR (\#FA9550-17-1-0327) for supporting this work. RD carried out this work at the USC Viterbi School of Engineering as part of the INDO - U.S. Science and Technology Forum (IUSSTF).

\newpage
\bibliographystyle{splncs03}
\bibliography{ref}

\end{document}